\documentclass[11pt]{article}
\usepackage{epsfig}
\usepackage{amsmath}
\usepackage{latexsym}

\newtheorem{lemma}{Lemma}

\newtheorem{theorem}[lemma]{Theorem}

\newcommand{\qed}{\hspace{1em}\hfill\rule{.5em}{.5em}}
\newenvironment{proof-of}[1]{\paragraph{Proof of {#1}:}\ 
  }{\mbox{}\hfill\qed\medskip}

\newcommand\eps\varepsilon

\newcommand\drop[1]{}

\newcommand\href[2]{#2}

\begin{document}

\title{Secondary Indexing in One Dimension: \\ Beyond B-trees and
  Bitmap Indexes\thanks{The main ideas for this work were conceived during Dagstuhl seminar No.\ 08081 on Data Structures, 2008.}}

\author{
Rasmus Pagh\thanks{IT University of Copenhagen, Denmark.
{Email: pagh@itu.dk}}
\and 
S.~Srinivasa Rao\thanks{MADALGO (Center for Massive Data Algorithmics, a center of the Danish National Research Foundation), Aarhus University, Denmark.
{Email: ssrao@madalgo.au.dk}}
}

\maketitle

\begin{abstract}
Let $\Sigma$ be a finite, ordered alphabet, and let ${\bf
x}=x_1x_2\dots x_n\in \Sigma^n$. A {\em secondary index\/} for ${\bf
x}$ answers alphabet range queries of the form: Given a range
$[a_l,a_r]\subseteq\Sigma$, return the set $I_{[a_l;a_r]}=\{i \;|\;
x_i \in [a_l; a_r] \}$. Secondary indexes are heavily used in
relational databases and scientific data analysis. It is well-known
that the obvious solution, storing a dictionary for the set $\bigcup_i
\{x_i\}$ with a position set associated with each character, does not
always give optimal query time. In this paper we give the first
theoretically optimal data structure for the secondary indexing
problem. In the I/O model, the amount of data read when answering a
query is within a constant factor of the minimum space needed to
represent $I_{[a_l;a_r]}$, assuming that the size of internal memory
is $(|\Sigma| \lg n)^\delta$ blocks, for some constant $\delta > 0$.
The space usage of the data structure is $O(n\lg |\Sigma|)$ bits in
the worst case, and we further show how to bound the size of the data
structure in terms of the $0$th order entropy of~${\bf x}$.
We show how to support updates achieving various time-space trade-offs.

We also consider an approximate version of the basic secondary
indexing problem where a query reports a superset of $I_{[a_l;a_r]}$
containing each element not in $I_{[a_l;a_r]}$ with probability at
most $\varepsilon$, where $\varepsilon > 0$ is the false positive
probability. For this problem the amount of data that needs to be read
by the query algorithm is reduced to $O(|I_{[a_l;a_r]}|
\lg(1/\varepsilon))$ bits.
\end{abstract}

\newpage


\section{Introduction}

Indexing capability is a vital part of database systems, and hundreds
of indexing methods exist. In this paper we consider indexes that
store a multiset of keys from an ordered set $\Sigma$, where each key
has some associated data. The goal is to support {\em range queries}
finding all keys (and associated data) in a given interval. In
databases a distinction is made between {\em primary\/} indexes, where
the data associated with each key is stored in the index itself, and
{\em secondary\/} indexes where the index provides references to the
associated data, which is stored in a way not controlled by the
index. The distinction is especially important in the I/O model, where
the time to read the references is usually much smaller than the time
to retrieve the associated data. This is because the associated data
is, in general, unlikely to be located in a small number of disk
blocks.

At first glance it would seem that the performance of secondary
indexes is not too important in the case where the set of returned
references is a large fraction of all data (in database jargon, where
the {\em selectivity\/} is low), since then the time to read the
associated data will dominate. However, it is common to use several
secondary indexes in conjunction. For example, in a database of people
we may want to find all married men of age 33. This can be done by
combining information found in secondary indexes for the attributes
specifying marital status, sex, and age. Only associated data matching
all three conditions needs to be returned. This means that the time
spent by the secondary indexes may be dominant, even when retrieving
the associated data is taken into account. This way of using secondary
indexes, often referred to as {\em RID intersection}, is particularly
common in On-Line Analytical Processing (OLAP) systems, information
retrieval, and scientific data analysis~(see
e.g.~\cite{rid-intersect,multires-bitmap,WAH} for details).

From a worst case query time perspective it would seem better to
support queries like the above using a data structure for orthogonal
range queries in three dimensions, e.g., a range tree. This would
ensure good query performance in terms of data size and result
size. However, when the number of dimensions is more than a small
constant (say, 3) known range reporting data structures either:
\begin{itemize}
\item Use excessive space (e.g., {\em range trees\/}~\cite[Section
5.3]{geombook} have space usage that grows with $(\lg n)^{d-1}$), or
\item Have no (provably) good worst-case performance (e.g., {\em
kD-trees\/}~\cite[Section 5.2]{geombook} have worst case query time
$n^{1-1/d}$).
\end{itemize}
This is one reason why it is common to perform multi-dimensional range
queries by intersection of the set of matching points in each
dimension, as explained above. Another reason is that one-dimensional
search structures are usually simpler and have lower constant factors
than multi-dimensional data structures.

Finally, using a collection of one-dimensional search structures
allows answering queries that are more general than orthogonal range
queries. Examples are approximate range searches (``find points that
are in the range in at least $d_1$ out of $d$ dimensions'') and
partial match queries (``find points that match range conditions in
$d_1$ given dimensions, where $d_1\ll d$''). For many of these problems,
all known solutions (for high dimensions) are not much better than the
brute force solutions (either represent all answers, or let queries
read most of the data).





\subsection{Problem definition}

Let $\Sigma$ be a finite, ordered alphabet, and let ${\bf
x}=x_1x_2\dots x_n\in \Sigma^n$. For a set $C\subseteq \Sigma$ we
define $I_{C}({\bf x})=\{i \;|\; x_i \in C\}$. When the string ${\bf
x}$ is understood we omit it. We formalize the {\em secondary indexing
problem\/} as follows: Let ${\bf x}=x_1x_2\dots x_n\in \Sigma^n$. A
secondary index for ${\bf x}$ answers alphabet range queries that,
given $a_l,a_r\in\Sigma$, return the set $I_{[a_l;a_r]}({\bf x})$. We
let $z=|I_{[a_l;a_r]}|$ and $\sigma=|\Sigma|$. Without loss of
generality we assume $\sigma \leq n$ (if it is larger, use a
dictionary to map to a smaller alphabet). In this paper we will
consider data structures that output the set in compressed format,
using $O\left(\lg\tbinom{n}{z}\right)$ bits. The size of the data
structure can be expressed either in terms of $n$ and $\sigma$, or more
generally in terms of the $0$th order entropy of ${\bf x}$.

In the {\em semi-dynamic\/} version we allow insertions of a new character at the end of ${\bf x}$. In the {\em fully dynamic\/} version we allow changing the character in a given position. We discuss in Section~\ref{sec:dynamic} how this is enough to also handle deletions.


We also consider a generalization of the secondary indexing problem. In the {\em approximate secondary indexing problem\/}  with parameter $\varepsilon \geq 0$ (Section~\ref{sec:approx}) the result of queries should be a set $\hat{I}_{[a_l;a_r]}\supseteq I_{[a_l;a_r]}$ such that for every $i\not\in I_{[a_l;a_r]}$, $\Pr[i\in \hat{I}_{[a_l;a_r]}]\leq \varepsilon$. The motivation for this problem is that it may be enough to filter away almost all points in the $d$-dimensional range query application. If a point is inside the range in $k$ dimensions, the probability that it will be reported by all $d$ approximate range queries is at most $\varepsilon^{d-k}$. False positives can be filtered away when accessing the associated data, assuming that the $d$ keys are stored with the associated data (which is typically the case in database applications).


\subsection{Previous work}

If $\Sigma$ has constant size, an optimal secondary index (up to
constant factors) is to store for every $a\in\Sigma$ a {\em bitmap
index\/} for the set $I_{\{a\}}$, i.e., the bit string ${\bf
x}_a\in\{0,1\}^n$ where there is a $1$ in position $i$ if and only if
$x_i=a$. A range query can be answered by simply reading the bitmap of
each character in the range.

If $\Sigma$ is large it is clear that some bitmaps will be very
sparse, and hence storing an explicit bitmap for each character will
be inefficient in terms of space. This problem can be addressed by
compressing the each bitmap to a representation that uses close to the
information-theoretic minimum space for representing vectors of a
given sparsity. A bit string of length $n$ with $m\leq n/2$ $1$s
requires space at least $\lg\binom{n}{m} = m\lg(n/m) + \Theta(m)$
bits.\footnote{We use $\lg$ to denote the logarithm base 2.}  One
well-known optimal encoding (within a constant factor) is the run-length
encoding, where the length $x$ of each run of $0$s is encoded using a
gamma code~\cite{elias:75:gammacodes} using $2\lfloor\lg
(x+1)\rfloor+2$ bits. Even though the bitmaps are not independent,
compressing them independently gives a total size of the data
structure that is within a constant factor of the size of the original
string ${\bf x}$, i.e., $O(n\lg \sigma)$ bits. While gamma coding is
asymptotically optimal, compression schemes used in practice also take
into account the computational effort needed to compress and
uncompress~\cite{WAH}, with some reduction in worst-case compression
rate.\footnote{Some bitmap indexing schemes such as~\cite{WAH} claim space optimality, but this is in comparison with indexes that represent all data explicitly, using at least $\log n$ bits per key, which is not optimal for dense sets.} 
In this paper we focus on the number of bits read and written,
and hence use run-length encoding with gamma codes (or more generally,
any method that compresses to within a constant factor of minimum
size). More formally, we will analyze our data structure in the {\em
I/O model\/}~\cite{MR90k:68029} where the cost measure is the number
of memory blocks read and written. (Note that in this model the minimum amount
of data read is 1 block, i.e., we count block I/Os and not merely the amount of data read.) 
While recent studies~\cite{WAH, conf/icde/SinhaWWSS08} 
indicate that computation time can be a bottleneck when handling
(compressed) bitmap indexes, we find it likely that I/O is going to be a
future bottleneck, as the parallel processing power of CPUs is rapidly
increasing.

Performing a range query by reading the compressed bitmap for each
character in the range does not always give the best query time we
could hope for. For example, if each character occurs $n/\sigma$ times
and we make a range query of size $\ell$ the output is a set of size
$n\ell/\sigma$, which can be represented in $O(\frac{n \ell}{\sigma}
\lg(\sigma/\ell))$ bits. However, the total size of the individual
bitmaps is $\Theta(\frac{n \ell}{\sigma} \lg \sigma)$, meaning that we
are reading a factor $\Omega(\lg(\sigma)/\lg(\sigma/\ell))$ more bits
than the output size. When $\ell = \Omega(\sigma)$ this is a factor
$\Omega(\lg\sigma)$ from optimal.

There have been a number of papers trying to use some kind of
precomputation to allow faster range queries. Some of these, such as
{\em range encoding\/}~\cite{ONeil:1997:IQP} and {\em interval
encoding\/}~\cite{Chan:1998:BID,Chan:1999:EBE} use space $n
\sigma^{1-o(1)}$ bits~\cite{multires-bitmap}. In this paper we are
interested in schemes where the precomputed data structure uses space
close to the minimum possible for representing ${\bf x}$. A more
space-conscious approach to range queries is {\em binning\/}
(see~\cite{multires-bitmap}). In its simplest form the idea is to
divide $\Sigma$ into bins of $w$ characters and represent a compressed
bitmap for each bin corresponding to all occurrences of its
characters. This means that a range query where the range has size
$\ell$ can be answered by combining less than $\lfloor\ell/w\rfloor +
2w$ compressed bitmaps. Using this idea recursively one gets {\em
multi-resolution bitmap indexes\/}~\cite{multires-bitmap}. Though not
analyzed in~\cite{multires-bitmap} the worst-case space usage of such
an index, when each bitmap is optimally compressed, is
$\Theta(n\lg^2(\sigma)/\lg w)$ bits. Queries may in the worst case
require reading a factor $O(\lg w)$ more data than in the size of the
output. This means that there is a time-space trade-off, and one can
never simultaneously achieve optimal space for the data structure and
optimal query time. In fact, a more general scheme is discussed
in~\cite{multires-bitmap} that allows the bucket size to be different
on the various resolution levels, but even this does not seem to yield
any worst-case improvement.

\subsection{Our results}

In the static setting, we obtain a data structure that simultaneously
achieves two goals:

\begin{itemize}
\item Space usage that is within a constant factor of the size of the
string~${\bf x}$. In fact, the size of our data structure is within a
constant factor of the $0$th order entropy of~${\bf x}$, plus $O(n)$ bits. 
This is up to a factor $\Omega(\lg \sigma)$ less than the explicit representation of ${\bf x}$.
\item Time usage for range queries that is within a constant factor of
what would be needed to read the result, had it been precomputed. This is up to a factor $\Omega(\lg n)$ less than the time needed to read the explicit list of positions in the result.
\end{itemize}

Our result improves previous results, all of which exhibit a time-space
trade-off. 
We also show how to make our structure dynamic, which is something
that has not been achieved by earlier data structures. Depending on the
time allowed for updates, we achieve the same, or nearly the same, query time.
Finally, we show how to support Bloom filter-like approximate queries with
improved efficiency.

\noindent The main conceptual and technical contributions of the paper are:
\begin{itemize}
\item Formulation of the theoretical problem: Secondary indexing with worst-case optimal space and query time. This gives a unified view of secondary index performance, with B-trees and uncompressed bitmap indexes at the extremes.
\item A new multi-resolution bitmap indexing scheme that (ignoring constant factors) does not exhibit a trade-off between space and query time. Section~\ref{sec:nonuniform-index}.
\item A dynamization of the data structure (Section~\ref{sec:dynamic}). A component of this, which is of independent interest, is a dynamic, buffered bitmap index (Section~\ref{sec:bufferbitmap}). The dynamization is mainly a technical contribution, requiring the use of several known ideas.
\item An I/O efficient way of supporting approximate range queries. The set returned by such a query is rather large, but we show how it can be highly compressed such that the representation is significantly smaller than that of an exact result.
\end{itemize}

\subsection{More notation}

We define the {\em cardinality\/} of a bitmap $S$ to be the number of
$1$s in it.
Let $T$ denote the size of the
(optimally) compressed output in bits. It will be convenient for us to
measure the block size $B$ of the I/O model in {\em bits}, rather than
words. Similarly, $M$ will denote the size of internal memory in bits.
We also use the parameter $b$ to denote the block size in ``words'',
i.e., $b = \Theta(B/ \lg n)$, where $n$ is the size of the input. We
assume that $B \ge \lg n$, and also that $b \ge 2$.  Let $\Sigma = \{
a_1, a_2, \dots, a_{\sigma} \}$ with $a_{1} < a_{2} < \dots <
a_{\sigma}$.

\section{Our secondary index}

To simplify the description of the structure below, we assume $\sigma$
to be a power of $2$.  The structure and its analysis can be easily
modified to work for the general case.


\subsection{A suboptimal solution.}
As a warm-up we first describe a simpler structure that introduces
some of the ideas used later, but only gives a suboptimal result.
The data structure is a variant of multi-resolution bitmap indexes~\cite{multires-bitmap},
and is not space optimal, but it is a suitable stepping stone for the
optimal solution described in Section~\ref{sec:nonuniform-index}.

Consider the complete binary tree $U$ with $\sigma$ leaves
identified from left to right with the sequence $a_1, a_2, \dots,
a_{\sigma}$.  With the leaf $a_i$ we associate the bitmap $I_{a_i}({\bf
x})$, and with each internal node $v$, we associate the bitmap
$I_{[a_l;a_r]}({\bf x})$ where $a_l$ and $a_r$ are the leftmost and
rightmost leaves below $v$, respectively.

Let $v_{1}, v_{2}, \dots, v_{2^{j}}$ be the nodes, in left-to-right
order, at level $j$ (the root being at level $1$), and let $n_{i}$ be the
cardinality of the bitmap associated with the node $v_{i}$. We store
the compressed bitmaps of all the nodes at each level in their
left-to-right order. The space used by all the compressed bitmaps at
the $j$th level is $O\left(\sum_{i=1}^{2^j} \lg {n \choose
n_i}\right)$. This summation is maximized when each of the $n_{i}$s is
equal to $n/2^{j}$, and in this case the space used by the $j$th level
compressed bitmaps is $O\left(\sum_{i=1}^{2^j} \lg {n \choose
n/2^{j}}\right)$, which is $O(n j)$ bits. Hence the total space used
by all the levels is $\sum_{j=0}^{\lg \sigma} O(n j) = O(n \lg^{2}
\sigma)$ bits.  We store the bitmaps of all the nodes in their level
order (from top to bottom, and form left to right in each level).
For each node, we also store the position and length of its compressed
bitmap, which takes $O(\sigma \lg n)$ bits overall.

We store an array $A$ of length $\sigma + 1$ where $A[i]$ stores the
cardinality of the bitmap $I_{[a_1;a_i]}({\bf x})$, for
$i=1,\dots,\sigma$, and $A[0]=0$.  The cardinality of
$I_{[a_l;a_r]}({\bf x})$ is $z = A[r] - A[l-1]$.  If $z > n/2$,
then instead of computing the answer to the original query
$I_{[a_l;a_r]}({\bf x})$, we compute the answers to the two queries
$I_{[a_1;a_{l-1}]}({\bf x})$ and $I_{[a_{r+1};a_{\sigma}]}({\bf x})$,
and return their union (which is the complement of the query
result). We now show how to answer a query for $I_{[a_l;a_r]}({\bf
x})$ assuming $z \le n/2$.

To answer an alphabet range query, $I_{[a_l;a_r]}({\bf x})$, we first
observe that any consecutive range of leaves can be covered by the
disjoint union of $O(\lg \sigma)$ subtrees of $U$ (by taking the
maximal subtrees for which all leaves are within the range -- at each
level, there will be at most two maximal subtrees whose leaves are
within the range). We compute the compressed bitmap of their union by
merging the bitmaps. Assuming that the size $M$ of internal memory is
at least $B\log \sigma$ this can be done in a single pass.

Since the cardinalities of the bitmaps associated with subtrees
decrease by a factor of $2$ as we go down by one level in the tree, we
can argue that the sum of the sizes of these $O(\lg \sigma)$ subtrees
(where there are at most two subtrees at each level) is at most $4$
times the size of a subtree that is present at the highest level.
Also, the length of the compressed bitmap associated with a subtree
that is present at the highest level is a lower bound on $T$, since
$z\le n/2$. Thus the overall number of I/Os is $O(\lg\sigma)$ plus an
$O(T/B)$ term for reading the bitmaps.

The result of the warm-up case is the following theorem, which we improve
later.

\begin{theorem}
\label{thm:uniform}
  A string ${\bf x}=x_1x_2\dots x_n\in \Sigma^n$ over a finite
  alphabet $\Sigma$ of size~$\sigma$ can be stored using $O(n
  \lg^2 \sigma)$ bits\footnote{The space bound can be improved to 
  $O(n \lg \sigma + \sigma \lg^2 n)$ bits while improving the query time 
  to $O(T/B + \lg\lg \sigma + \lg_b \sigma)$ I/Os by using the ideas 
  described in Section~\ref{sec:nonuniform-index}.} 
  such that range queries can be answered in $O(T/B+\lg\sigma)$ I/Os.
\end{theorem}

\subsection{Optimal space structure.}\label{sec:nonuniform-index}

We now describe how to extend the above scheme to the case of
non-uniform distribution, and at the same time reducing the space to optimal
(assuming $\sigma \ll n$). 
For simplicity we assume that no character
has more than $n/2$ occurrences. If this is not the case we may expand
the alphabet and substitute half of the occurrences of the most common
character with a new character, increasing the $0$th order entropy by
$O(n)$ bits. A main tool is to use a ``weight balanced tree'' $W$ on
the multi-set of characters occurring in ${\bf x}$, instead of the
complete binary tree $U$ on the alphabet (used in the case of uniform
distribution). Each of the $n$ characters is associated with its
position in ${\bf x}$; their ordering in the tree is determined
primarily by the order on $\Sigma$, secondarily by the ordering of
positions.

Our starting point is a weight-balanced B-tree
from~\cite{Arge:2003:OEM} with constant maximum degree. As in other
B-tree variants, all leaves are at the same distance from the root.
The essential property we need is that in a weight-balanced B-tree
with branching parameter $c>4$ (a constant) and leaf parameter $1$ we
can efficiently maintain that the {\em weight\/} of a node $v$ (number
of leaves in the subtree below $v$) at level $i$ {\em from the
bottom\/} is between $\tfrac{1}{2}c^i$ and $2c^i$. This implies that
the maximum degree is $4c$ and the depth is $O(\lg n)$. The weight of
a node is stored with the node. Note that the weight of a node at
level $i$ {\em from the top\/} is $\Theta(n/c^i)$ (this is the bound
we will actually use).



With each internal node $v$ of the tree, we associate a compressed
representation of a bitmap $S_v$ of length $n$ where $S_v[i] = 1$ if
the character in position $i$ is in the subtree rooted at $v$, and
$S_v[i]=0$ otherwise.  We now prune this tree by removing all the
children of an internal node $v$ if all leaves below $v$ contain the
same character.  In this pruned tree, each character appears at most
$8c$ times at each level as a leaf (if it appears more than $8c$ times
as a leaf at a particular level, then since all these leaves must be
adjacent, a subset of them will be the only children of their parent
and hence should have been removed by the pruning procedure as their
parent would be associated with a single character). Thus the total
number of leaves, and hence the total number of nodes in the pruned tree
is $O(\sigma \lg n)$, for constant~$c$.  We use $W$ to denote this
pruned tree.  We define the weight of a node in the pruned tree to be
the cardinality of the bitmap associated with it (which is same as its
``weight'' in the tree before pruning).

\paragraph{A naive upper bound.}
Since the sum of the cardinalities of all bitmaps stored with nodes at
any particular level is at most $n$ and since each bitmap of a node at
level $i$ has cardinality $\Theta(n/c^{i})$, the compressed bitmap of
a node at level $i$ takes $\Theta(n i / c^{i})$ bits (recall that $c$
is constant).  Since there are $O(c^i)$ nodes at level $i$ the total
space at this level is $O(ni)$ bits, and because the height of the
tree is $O(\lg n)$, the overall space used by all the levels is $O(n
\lg^{2} n)$ bits.
The tree structure (including pointers to the bitmaps at each node,
but not including the bitmaps themselves) is laid out on the disk such
that any root-to-leaf path can be traversed using $O(\lg_{b} n)$ I/Os.
More specifically, starting from the root, we store the top $d =
\Theta(\lg b)$ levels in a block with pointers to each of the subtrees
at level $d+1$. Each of the subtrees at level $d+1$ are recursively
stored in the same fashion. We merge the blocks so that no block is
more than half empty. Thus the total space is within a constant factor
of the space needed to store the tree without ``blocking''. Since we
can traverse $\Theta(\lg b)$ levels in any root-to-leaf path using one
I/O, any root-to-leaf path can be traversed using $O(\lg_b n)$ I/Os.
The compressed bitmaps of all nodes are stored in level order.

A range query $I_{[a_l;a_r]}({\bf x})$ can be partitioned into $O(\lg
n)$ subtrees where there are only a constant number of subtrees at
each level.  These subtrees can be identified by traversing the tree
top down to find the leftmost and rightmost leaves associated with the
characters $a_{l}$ and $a_{r}$ respectively, using $O(\lg_{b} n)$
I/Os. More specifically, starting from the root, at each level we go
to the leftmost (rightmost) child of the current node which is
associated with $a_{l}$ ($a_{r}$), while including subtrees rooted at
all the right (left) siblings of the child to the list of subtrees to
be merged.  If $z$ is the cardinality of the query result, then there
are $O(1)$ subtrees of weight between $z$ and $z/c$, $O(1)$ subtrees
of size between $z/c$ and $z/c^2$, $\dots$, $O(1)$ subtrees of size
$1$. Thus the total size of all the compressed bitmaps we need to read
(to compute their union) is at most $O(1) ( z \lg (n/z) + (z/c) \lg
(cn/z) + \dots + \lg n ) = O(z \lg (n/z))$, which is asymptotically
optimal. For each compressed bitmap read we may use up to $2$ I/Os to read blocks
that do not contain $B$ bits of the compressed bitmap. That is, we waste
 $O(\lg n)$ I/Os compared to the smallest possible number of I/Os required
 to read the compressed bitmaps.

\paragraph{Space improvement.} 
We now show how to improve the space of the above scheme to $O(n
H_{0})$ bits, where $H_{0}$ is the $0$th order entropy of the string
${\bf x}$, while retaining the query time. The main idea is to store
bitmaps only on a few levels of the tree explicitly instead of storing
all of them.  More specifically, if $h$ is the height of the tree
$W$, we store the $O(\lg h)$ levels numbered $1, 2, 4, 8, \dots$
(from the top), and also store all the leaves explicitly. We refer to the levels 
that are stored explicitly (including the leaf level) as
the {\em materialized levels}. In addition, we store the structure of the entire
tree~$W$, without removing any levels. 

We now analyze the space usage. Consider an instance of a character
that is stored at level $i$ in the tree. It contributes to the space
usage of one bitmap index at each of levels $1,2,4,\dots,2^{\lfloor
\lg i \rfloor}$, which means a total of $O(i)$ bits. In other words,
the total space is within a constant factor of the space needed for
the bitmap indexes stored at the leaves.

Now consider the leaves corresponding to a single character
$a\in\Sigma$ with $z_a$ occurrences. As noted above, at any level
there are at most $8c$ leaves associated with $a$. Further, the
heaviest leaf associated with $a$ has weight $\Theta(z_a/c)$ (because
a constant fraction of the total weight must come from the uppermost
level). As argued earlier, the exponential decrease in weight down the
tree means that the space usage is dominated by the uppermost level,
which uses space $O(z_a\lg(n/z_a))$. Summing over all characters we
get $\sum_{a\in\Sigma} O(z_a\lg(n/z_a)) = O(n H_0)$ bits, where $H_0$
is the $0$th order entropy of the string ${\bf x}$. In addition, the
tree has $O(\sigma \lg n)$ nodes, each of which stores a pointer to
the bitmap corresponding to that node. Since each pointer can be
represented using $O(\lg n)$ bits, the total space used by the tree
structure is $O(\sigma \lg^{2} n)$ bits.

To answer a query we use the same algorithm as in the ``naive upper
bound'', except that when we need a bitmap index that is not
explicitly stored, it is computed by merging the bitmaps stored with
all the nearest descendants that are in the materialized level
immediately below. The weight of a node is used to figure out the
total cardinality of the bitmaps to be read from the next materialized level.
We store the bitmaps of all the internal nodes at each materialized
level by concatenating them in their left-to-right order. The set of all
bitmaps to be read at a level form two consecutive chunks in the 
concatenated sequence of bitmaps. So at each
materialized level, and at the leaf level, $O(1)$ I/Os are wasted
reading the data that is not needed to form the answer.
Assuming that $M=B(\sigma\lg n)^{\Omega(1)}$ we can merge these
$O(\sigma \lg n)$ bitmaps in $O(1)$ passes (the weights of the nodes
can be used to find the starting positions of the compressed bitmaps), 
meaning that the number of I/Os is within a constant factor of the I/Os 
needed to read the individual bitmaps.

We now analyze the query time. Searching the tree $W$ to find all 
the relevant subtrees to be merged requires $O(\lg_{b} n)$ I/Os. From 
these subtrees, one can compute the starting positions and the lengths
of all the bitmaps to be read from all the materialized levels (without 
any additional I/Os). The space needed to represent the
compressed bitmap of a node at level $i$ in the tree is bounded by a
constant factor (two) times the space used by its lowest ancestor that
is stored explicitly (i.e., the ancestor of the node at level
$2^{\lfloor \lg i \rfloor}$). This means that the number of bits we
need to read is within a constant factor of the case where all bitmaps
are explicitly stored, which is the case analyzed in the ``naive upper
bound'' above. In addition, we waste $O(1)$ I/Os in reading the
relevant bitmaps at each materialized level, and hence overall
$O(\lg\lg n)$ I/Os are wasted. 

%

\begin{theorem}
\label{thm:general}
  A string ${\bf x}=x_1x_2\dots x_n\in \Sigma^n$ over a finite
  alphabet $\Sigma$ of size~$\sigma$ can be stored using $O(n H_{0} + n +
  \sigma \lg^{2} n)$
  bits, where $H_{0}$ is the $0$th order entropy of~${\bf x}$, such
  that range queries can be answered in $O(z \lg (n/z)/B + \lg_{b} n + \lg\lg n)$
  I/Os, where~$z$ is the cardinality of the answer to the query,
  assuming $M=B(\sigma\lg n)^{\Omega(1)}$.
\end{theorem}


\section{Approximate queries}\label{sec:approx}

We now consider how to reduce the query time by allowing queries to
return false positives, in the spirit of Bloom
filters~\cite{Bloom:1970:STT,MR80h:68037}. More precisely, a query
reports a set $\hat{I}_{[a_l;a_r]} \supseteq I_{[a_l;a_r]}$, where for
each $i\not\in I_{[a_l;a_r]}$ the probability that $i\in
\hat{I}_{[a_l;a_r]}$ is at most $\varepsilon$. The parameter
$\varepsilon$ is supplied as an argument to the query
algorithm. Approximate secondary indexing was recently considered by
Apaydin et al.~\cite{conf/vldb/ApaydinCFT06}, but their query
algorithm is optimized for a RAM model meaning that it has many random
accesses and thus poor performance in the I/O model.

We use the technique of Carter el al.~\cite{MR80h:68037}, as further
developed by Bille et al.~\cite{intersect}, for converting the problem
of storing a set with $\varepsilon$ false positive rate to the problem
of storing exactly a set within a smaller universe.  Specifically,
whenever the exact data structure described in
Section~\ref{sec:nonuniform-index} stores a set of positions
$S\subseteq [n]$, the approximate data structure additionally stores a
sequence of $k=\lfloor \lg \lg n\rfloor$ hashed sets
$h_1(S),\dots,h_k(S)$, where $h_j: U\rightarrow [2^{2^j}]$ is a
function chosen at random from a universal family. The same $k$
functions are used in each node, and we group the sets according to
what hash function was used. At a node where a set $I$ is stored, the
hashed sets occupy $O(\sum_{j=1}^k \lg\binom{2^{2^j}}{|I|}) =
O(\lg\binom{n}{|I|})$ bits. This means that the total space needed to
store the hashed sets, as compressed bitmaps, is dominated by the
space needed to store $S$.

When processing a query for the interval $[a_l;a_r]$ the first step is
to compute the size $z$ of the result $I_{[a_l;a_r]}$. This can be
done efficiently using the weight-balanced B-tree. Then we choose $j$
as the smallest integer such that $2^{2^j} > z/\varepsilon$.  If $j>k$
we cannot save anything by returning an approximate result, so we
answer the query exactly as described in
Section~\ref{sec:nonuniform-index}. Otherwise we compute
$h_j(I_{[a_l;a_r]})$ by taking the union of the $j$th hashed sets
(rather than the union of the position sets themselves). By the
analysis in Section~\ref{sec:nonuniform-index} the number of bits read
is $O(\lg\binom{2^{2^j}}{z})$, which is $O(z\lg(1/\varepsilon))$ by
our choice of $j$.

Finally, we let $\hat{I}_{[a_l;a_r]}$ be the preimage
$h_j^{-1}(h_j(I_{[a_l;a_r]}))$ of the hashed result. For many common
universal families the preimage can be computed with a small
effort. Note that we do not want to output the preimage (it is quite
large), but only to generate it without using any further I/Os. In the context of
high-dimensional range queries it is also important that we can efficiently
compute the intersection of several approximate query results, but this is easy:
Simply compute the preimage of the intersection. 

We describe a well-known and particularly attractive universal family. 
Split a number $i\in [n]$ into two parts $(i_1,i_2)$ where $i_2$
is the $2^j$ least significant bits of $i$ and $i_1$ is the $\lceil
\lg(n+1) \rceil - 2^j$ most significant bits of $i$. Then take any
universal family $\mathcal{H}$ mapping to $[2^j]$, pick
$g_j\in\mathcal{H}$ uniformly at random and let
$h_j(i_1,i_2)=g_j(i_1)\oplus i_2$ where $\oplus$ denotes bitwise
exclusive or. The family from which $h_j$ is chosen can easily be seen
to be universal. Then the set of indices $(i_1,i_2)$ that map to $s\in
\{0,1\}^{2^j}$ is $h_j^{-1}(s) = \{ (i_1,s\oplus g_j(i_1)) \; | \;
i_1=0,1,2,\dots \}$.

To argue that the desired error bound is met, consider $i\not\in
I_{[a_l;a_r]}$. By universality, the probability that $h_j(i)\in
h_j(I_{[a_l;a_r]})$ is at most $z/2^j \leq \varepsilon$. Since $i\in
\hat{I}_{[a_l;a_r]}$ if and only if $h_j(i)\in h_j(I_{[a_l;a_r]})$
this completes the argument. We get the following approximate
variant of Theorem~\ref{thm:general}.

\begin{theorem}\label{thm:approximate}
  A string ${\bf x}=x_1x_2\dots x_n\in \Sigma^n$ over a finite
  alphabet $\Sigma$ of size~$\sigma$ can be stored using $O(n H_{0} + n +
  \sigma \lg^{2} n)$
  bits, where $H_{0}$ is the $0$th order entropy of~${\bf x}$, such
  that approximate range queries with false positive probability $\varepsilon$ 
  can be answered in $O(z \lg (1/\varepsilon)/B + \lg_{b} n + \lg\lg n)$
  I/Os, where~$z$ is the cardinality of the answer to the query,
  assuming $M=B(\sigma\lg n)^{\Omega(1)}$. The I/O bound captures the
  time for generating the result, but not for outputting it.
\end{theorem}

As shown by Carter et al.~\cite{MR80h:68037} the space needed to represent
a set of size $z$ approximately, with false positive probability $\varepsilon$, is 
$O(z \lg (1/\varepsilon))$ bits, so the query time is optimal whenever $z$ is not small (e.g., when $z>\lg n$).


\section{Supporting updates}\label{sec:dynamic}

Given a string ${\bf x}$ over the alphabet $\Sigma$, we consider the
following update operations, for some $\alpha \in \Sigma$:
\begin{itemize}
\item $append({\bf x}, \alpha)$: append the character $\alpha$ at the
  end of ${\bf x}$, and
\item $change({\bf x},i,\alpha)$: change the $i$th character of ${\bf
  x}$ to $\alpha$.
\end{itemize}

We note that deletions are indirectly supported through these
operations: Extend the alphabet with a new character $\infty$ that is
newer matched by a range query. Deleting a character can be done by
simply changing it to~$\infty$. If deletion markers are similarly used
in the table being indexed this yields the desired semantics: The
positions of characters do not change when deletions are performed. It
is, however, simple to extend this to the more natural semantics where
character positions are always relative to the current string:
Maintain a B-tree over the deleted positions with subtree sizes
maintained in all nodes --- this allows translating positions back and
forth between the two systems using $O(\log_b n)$ I/Os, and space
$O(n)$ bits (positions in leaf nodes should be efficiently encoded,
e.g., using gamma-coded differences). If the number of deleted
characters exceeds a constant fraction of all characters, global
rebuilding is performed to reduce the space.

\subsection{Semi-dynamic version}\label{sec:semidynamic}

We first consider only updates that append characters to the end
of the string. This is motivated by the fact that OLAP and scientific data, for
which bitmap indexes have been shown to be very effective, are
typically {\em read and append only}~\cite{multires-bitmap}.
We describe two ways to modify the weight balanced B-tree structure
used in Section~\ref{sec:nonuniform-index} to support $append$,
achieving different time-space trade-offs.

A straightforward way of supporting updates to the structure described in
Section~\ref{sec:nonuniform-index} is to perform the update on all the
bitmaps that are affected by it. Since one bitmap in each materialized
level (namely the one corresponding to the last occurrence of that
character) will be affected by an update, we need to update $O(\lg\lg
n)$ bitmaps. This can be done efficiently by maintaining an array of
size $O(\lg\lg n)$ for each character $a \in \Sigma$. The $i$th entry
in the array corresponding to $a$ stores a pointer to the disk block
containing the last occurrence of $a$ among all bitmaps at the $i$th
materialized level. We also maintain back pointers so as to update 
these arrays efficiently whenever the blocks pointed to by the array
are reorganized. This requires an additional $O(\sigma \lg\lg n)$
pointers, using $O(\sigma \lg n \lg\lg n)$ bits.

After performing an update, if the weight-balancing condition at a
node is violated, then the subtree rooted at the parent of that node
is rebuilt to maintain the weight-balancing condition.  More formally, let $v$ be a
node at level $i$ from the top, and let $u$ be the parent of $v$. Let
$v$ be the highest level node at which the weight-balancing condition
is violated after an update. To maintain the weight-balancing
condition, we re-build the subtree rooted at~$u$, and recompute the
new bitmaps associated with all the nodes in the subtree. This can be
done bottom-up in level order. The bitmaps associated with leaves need not
be recomputed. The bitmaps associated with all the nodes at a materialized 
can be computed by merging the bitmaps of all the descendants at the
materialized level immediately below. Computing the compressed bitmap of 
a node by merging the compressed bitmaps of all its descendants 
(in the materialized level immediately below) requires $O(1)$ passes
assuming $M = B \sigma^{\Omega(1)}$, where each pass scans all the 
bitmaps that need to be merged exactly once.
Also the size of all the bitmaps at a any level within a given subtree 
is at most the size of all the bitmaps at the leaves within the subtree.
Thus overall, the number of I/Os needed to compute the bitmaps of
all the nodes in the subtree is at most $\lg\lg n$ (the number of levels) times the 
number of I/Os needed to scan the bitmaps of all the leaves in the subtree. 
As the weight of $u$ is $\Theta(n/c^{i-1})$,
the size of the bitmaps associated with all the leaves below $u$  
is $O((n/c^{i-1}) \lg (c^{i-1}))$ bits. Hence the cost of scanning the bitmaps 
associated with all the leaves in the subtree rooted at $u$ is 
$$(1/B)(\lg\lg n - i) O((n/c^{i-1}) \lg (c^{i-1})) \text{ I/Os.}$$ 
The total cost of rebalancing $u$ can be charged to the
$\Theta(n/c^i)$ updates performed in the subtree rooted at $v$ (which
caused the violation in the weight-balancing condition at~$v$), making
the amortized cost of rebalancing $O(\frac{(\lg\lg n)^{3}}{B})$ which is $O(1/b)$ I/Os.

%
Thus this structure extends the structure of Theorem~\ref{thm:general}
by supporting updates in $O(\lg\lg n)$ amortized I/Os (while retaining
the space and query bounds).

\begin{theorem}
A string ${\bf x}=x_1x_2\dots x_n\in \Sigma^n$ over a finite alphabet
$\Sigma$ of size~$\sigma$ can be stored using $O(n H_0 + \sigma
\lg^{2} n)$ bits to support range queries in $O(z \lg (n/z)/B + \lg\lg
n + \lg_{b} n)$ I/Os, and $append$ in amortized $O(\lg\lg n)$ I/Os.
\end{theorem}


\subsubsection{Trading off space for faster updates.} 

To get faster updates, the main idea is to use buffers with the
internal nodes to store the updates, similar to buffer trees
\cite{Arge/03} and buffered B-trees~\cite{Brodal03a,Graefe:2006:BTI},
instead of performing them right away. With each internal node of the
tree $W$, we associate a buffer of size $B$ bits. (Note that we also
associate buffers with nodes that have no explicitly stored bitmap.)
The total space used by all the buffers is $O(B \sigma \lg n)$ bits,
as $W$ contains $O(\sigma \lg n)$ nodes, each of which is associated
with a buffer of $B$ bits.

We now describe how to support $append$. To perform $append(\alpha,
{\bf x})$, we first insert the instruction into the buffer of the
root, which is always kept in the internal memory. When the buffer at
a node $u$ becomes full, we find a child $v$ of $u$ on which at least
a (fixed) constant fraction of these updates have to be performed.
Since the degree of each node in $W$ is bounded by~$4c$ (a constant),
such a node always exists. If node $u$ is stored explicitly, then we
perform these updates on the bitmap associated with $u$. We then
delete those updates from the buffer at $u$ and insert them into the
buffer at node~$v$. Since the buffer size is $B$ bits, $\Theta(B/\lg
n) = \Theta(b)$ updates are moved from $u$ to $v$. Thus the amortized
number of I/Os required to perform an update on all the nodes on a
root-to-leaf path is $O(\frac{\lg n}{b})$.  The amortized cost of
rebalancing the weight-balanced tree can be shown to be
$O(\frac{1}{b})$ I/Os as mentioned before (the amortized cost
of reading all the buffers in the subtree being rebalanced is
negligible).

To answer a query, apart from performing the query algorithm described
in Section~\ref{sec:nonuniform-index}, we also need to read each of
the buffers associated with all nodes that could potentially contain
updates that are part of the answer to the query. 
Also, whenever we read the bitmap stored at a node, we do not need to
look at the buffers associated with any of its descendants, as all the
updates stored in the buffers at the descendants of a node have
already been performed on the bitmap associated with that node. Hence
the number of buffers we need to read to answer a query is only $O(\lg
n)$. Thus we have

\begin{theorem}
A string ${\bf x}=x_1x_2\dots x_n\in \Sigma^n$ over a finite alphabet
$\Sigma$ of size~$\sigma$ can be stored using $O(n H_0 + \sigma \lg n
(B + \lg n))$ bits to support range queries in $O(z \lg (n/z)/B + \lg
n)$ I/Os, and $append$ in amortized $O(\frac{\lg n}{b})$ I/Os.
\end{theorem}


\subsection{Buffered compressed bitmap index}\label{sec:bufferbitmap}
Let ${\bf x} = x_1x_2\dots x_n$ be a string over a finite alphabet
$\Sigma$. Given $\alpha \in \Sigma$, a {\em point query} returns (a
compressed representation of) the set $I_{\alpha}({\bf x}) = \{i | x_i
= \alpha \}$.  In this section, we first develop a structure that
dynamizes the standard bitmap index while supporting point queries
efficiently. We then use this structure (in
Section~\ref{sec:fullydynamic}) to obtain a fully dynamic structure
supporting alphabet range queries efficiently.  The structure
described below supports point queries in $O(\lg n + T/B)$ I/Os where
$T$ is the total size of the output, and updates in $O(\frac{\lg
n}{b})$ I/Os.

The main idea is to store the compressed bitmaps of all the characters
in a buffer tree.  First, each bitmap is represented in compressed
form as a list of positions of $1$s, in increasing order in the
bitmap, and this list is stored in a sequence of blocks. The first
position in each block is stored as an absolute value, and all the
others are stored relative to the previous position (i.e., the gaps
are encoded) using gamma codes. We concatenate the representations of
all bitmaps together and store them as a sequence of blocks.

Let $s = O(n H_0/B)$ be the number of blocks used by the compressed
bitmaps of all the characters.  Assuming $B \ge 4 \lg n$, we can bound
the number of blocks used by the representations of all the blocks in
terms of $s$. Store the original compressed bitmaps of the given
materialized level with blocks of $B/2$ bits each, and hence an
overall $2s$ blocks. By increasing the block size to $B/2 + 2 \lg n$,
we can make the first gamma code in each block to be an absolute value
instead of the being relative to the previous position (using at most
$\lg n$ additional bits), and the last code to be completely contained
within the block in case it is split between two adjacent blocks (by
using an additional $\lg n$ bits). Thus the new representation of the
blocks requires at most $2s$ blocks of size at most $B$ bits each.

With these blocks as the leaves, we construct a tree with branching
factor~$c$, for some fixed constant $c \ge 2$. With each internal node
of this tree, we associate a buffer of size $B$ bits that stores a set
of updates that are yet to be performed in one of the leaves in the
subtree rooted at that node.  Each non-leaf block also stores an
identifier for the first bitmap that is (partially) stored in the
subtree, to allow fast navigation to a particular bitmap.
%


An update is simply stored in the buffer corresponding to the root,
which is always kept in the internal memory. Whenever a buffer becomes
full, a constant fraction of the updates in that buffer are moved to
one of its children.  The amortized cost of updates can be shown to be
$O(\frac{\lg n}{b})$ I/Os as before. The space usage of this
structure is of the same order as the total space used by all the
individual compressed bitmaps, as the space usage is dominated by the
leaf level bitmaps.

A point query can be answered by following a root-to-leaf path in the
tree followed by reading the compressed bitmap stored in the
consecutive leaves starting at the end of the search path. In
addition, we also need to merge this compressed bitmap with all the
updates corresponding to the character, stored in the buffers. The
number of buffers storing updates corresponding to a given character
can be shown to be $O(T/B + \lg n)$ where $T$ is the size of the
compressed bitmap of the given character. Assuming $M \ge B \lg n$, we
can perform this merging in $O(1)$ passes. 
Thus we have

\begin{theorem}
A string ${\bf x}=x_1x_2\dots x_n\in \Sigma^n$ over a finite alphabet
$\Sigma$ of size~$\sigma$ can be stored using $O(n H_0)$ bits to
support point queries in $O(T/B + \lg n)$ I/Os, where $T$ is the size
of the answer, and updates in amortized $O(\frac{\lg n}{b})$ I/Os.
\end{theorem}


\subsection{Fully dynamic version} \label{sec:fullydynamic}
We first observe that all the bitmaps stored at any particular
materialized level of the optimal space structure described in
Section~\ref{sec:nonuniform-index} can be thought of as representing a
bitmap index over an alphabet containing one character corresponding
to each node in that level. Thus we can obtain a fully dynamic
secondary bitmap index by representing each of the materialized levels
as a buffered bitmap index.

More formally, we represent all the bitmaps at each materialized level
of the structure described in Section~\ref{sec:nonuniform-index} using
buffered bitmap index structure. The total space usage is $O(n H_{0} +
\sigma \lg^{2} n)$ bits (for representing all the bitmaps in all the
materialized levels, and for the tree structure). To perform an update
we perform it on each of the $O(\lg\lg n)$ materialized levels.  Thus
updates take amortized $O(\frac{\lg n \lg\lg n}{b})$ I/Os. An alphabet
range query can be decomposed into $O(1)$ point queries on each
materialized level, and can be answered using $O(\lg n \lg\lg n + z
\lg (n/z)/B)$ I/Os, where $z$ is the cardinality of the answer to the
range query.

\begin{theorem}
A string ${\bf x}=x_1x_2\dots x_n\in \Sigma^n$ over a finite
alphabet $\Sigma$ of size~$\sigma$ can be stored using  
$O(n H_0 + \sigma \lg^{2} n)$ bits to support range 
queries in $O(z \lg (n/z)/B + \lg n \lg\lg n)$ I/Os, 
and updates in amortized $O(\frac{\lg n \lg\lg n}{b})$ I/Os.
\end{theorem}

One can also achieve other trade-offs between space and operation
times by choosing to store all the levels of $W$ explicitly and
using buffers at the internal nodes.

%
%
%



\bibliographystyle{plain}
\bibliography{second}

\end{document}